\makeatletter
\providecommand*\input@path{}
\newcommand\addinputpath[1]{
  \expandafter\def\expandafter\input@path
  \expandafter{\input@path{#1}}}
\addinputpath{Figs/}
\addinputpath{Bios_photos/}
\makeatother

\documentclass[journal]{IEEEtran}
\usepackage{amsfonts}
\usepackage{amssymb}
\usepackage{amsmath}
\usepackage{diagbox}
\usepackage{algorithm}
\usepackage{multirow}
\usepackage{xcolor}
\usepackage{graphicx}
\usepackage{cite}
\usepackage{exscale}
\usepackage{relsize}
\usepackage{algpseudocode}
\usepackage{graphics}
\usepackage{mathrsfs}
\usepackage{siunitx}
\usepackage{threeparttable}
\usepackage{url}
\usepackage{booktabs}
\usepackage{lipsum}
\usepackage{array}

\DeclareMathOperator*{\argmin}{argmin}
\newcommand{\bm}[1]{\mbox{\boldmath{$#1$}}}

\newtheorem{Def}{Definition}
\newtheorem{Theo}{Theorem}
\newtheorem{Prop}{Proposition}

\ifCLASSOPTIONcompsoc
\usepackage[caption=false,font=normalsize,labelfont=sf,textfont=sf]{subfig}
\else
\usepackage[caption=false,font=footnotesize]{subfig}
\fi

\usepackage[T1]{fontenc}

\title{Sampling and Inference of Networked Dynamics using Log-Koopman Nonlinear Graph Fourier Transform}

\author{Zhuangkun Wei\textsuperscript{1}, Bin Li\textsuperscript{2}, Chengyao Sun\textsuperscript{3}, Weisi Guo\textsuperscript{1,3,4*}

\thanks{\textsuperscript{1}University of Warwick, UK. 
\textsuperscript{2}Beijing university of Posts and Telecommunications, Beijing, China.
\textsuperscript{3}Cranfield University, UK
\textsuperscript{4}The Alan Turing Institute, UK.
\textsuperscript{*}Corresponding Author: weisi.guo@cranfield.ac.uk. }}

\begin{document}

\maketitle

\begin{abstract}
Monitoring the networked dynamics via the subset of nodes is essential for a variety of scientific and operational purposes. When there is a lack of an explicit model and networked signal space, traditional observability analysis and non-convex methods are insufficient. 
Current data-driven Koopman linearization, although derives a linear evolution model for selected vector-valued observable of original state-space, may result in a large sampling set due to: (i) the large size of polynomial based observables ($O(N^2)$, $N$ number of nodes in network), and (ii) not factoring in the nonlinear dependency between observables.
In this work, to achieve linear scaling ($O(N)$) and a small set of sampling nodes, we propose to combine a novel Log-Koopman operator and nonlinear Graph Fourier Transform (NL-GFT) scheme. First, the Log-Koopman operator is able to reduce the size of observables by transforming multiplicative poly-observable to logarithm summation. Second, a nonlinear GFT concept and sampling theory are provided to exploit the nonlinear dependence of observables for observability analysis using Koopman evolution model. The results demonstrate that the proposed Log-Koopman NL-GFT scheme can (i) linearize unknown nonlinear dynamics using $O(N)$ observables, and (ii) achieve lower number of sampling nodes, compared with the state-of-the art polynomial Koopman based observability analysis.
\end{abstract}


\section{Introduction}
Many engineering, social, and biological complex systems consist of dynamical elements connected via a large-scale network. These include both explicit \cite{Krishnagopal17,Barzel13} and latent dynamics, spanning: urban structure \cite{Wilson08}, social networks \cite{8105893}, economics \cite{Bardoscia17}, engineering infrastructure \cite{Schafer18}, ecology \cite{Lu16}, biology clocks \cite{Hasegawa18PhyE}, epidemic spreading \cite{Scholtes14}, and organizational structure \cite{Ellinas17}. 

Collecting data on networks is important for a variety of scientific and practical reasons, ranging from scientific model development to Digital Twin informed maintenance \cite{8839864,9026801}. However, when the size of the network is large, as is the case for national infrastructure, gene regulatory networks, or social networks; effective monitoring through a small subset of critical nodes is essential. 
Optimal data collection (sampling) and inference in networked nonlinear dynamical systems is challenging. Current observability analysis and non-convex methods rely on either a state evolution model or the signal-space (e.g., sparsity or bandlimitedness - subject to a designed operator). Often, either the sparse property and/or the desirable operator to exploit this does not exist. 
As such, in the absence of explicit model and sparse signal-space to exploit, we tackle the challenge of how to characterize the dynamic evolution of the network, and how to use such evolution model for network sampling and signal recovery. We first leverage on the Koopman operator to derive a linearized evolution model of observable defined on original signal state. Next, we exploit the observable dependency to discover the optimal sampling points, and design signal recovery algorithm using nonliner GFT.

\subsection{Literature Review}

In classic topology-centric analysis, the influential nodes are often determined using eigen analysis resulting in wide measures such as PageRank centrality. However, when the dynamics are also important, the relationship between topological influence and cascade dynamics is unclear \cite{Gao16}. When nonlinear dynamics is coupled with complex networks, current methods fall into two categories. First, reduced order models (e.g. heterogeneous mean field around equilibrium conditions \cite{Moutsinas20}) cannot well approximate cascade transient dynamics. This means we can only understand the equilibrium conditions and the impact of perturbations.

\subsubsection{Model Driven}
In order to achieve transient behaviour understanding (also known as graph observability), a well-studied group of schemes is state-based reconstruction. Popular methods include convex optimisation \cite{4663892}, causal modeling \cite{7763882}, and observability analysis using linear evolution models (e.g., checking rank conditions of the linear model that maps initial state to all forward states, or maximizing energy of sampled states computed by model-relevant observability gramians) \cite{isufi2017observing, 7852369, zhang2017sensor, 7402218, 5411741}. These approaches all provide attractive performances on sampling node compression and state recovery accuracy, under the important premise of a known and linear/linearized underlying model. Their drawback is the inability to address the sampling and recovery challenges in the absence of dynamic equations.

\subsubsection{Data Driven}
Instead of relying on explicit dynamic equations, an alternative group resorts to the prior knowledge of the signal-space \cite{8839864}. The methods include sparsity and spectral analysis. For instance, the compressed sensing (CS) schemes \cite{Sidropoulos12, Ding17} selected the sampling nodes by analyzing the principal components. Graph sampling methods \cite{Chen15, anis2016efficient, Chen16, ortega2018graph, 7979523, isufi2017observing, 7979500, 7576646, 7934053, 7450694, 7480396, 8115204} determine a sampling node set for signals that belong to a sub-space (referred to as band-limited) of a Graph Fourier Transform (GFT) operator (e.g., Laplacian \cite{Chen15, anis2016efficient, Chen16, ortega2018graph, 7979500, 7576646, 7934053}, joint time-graph Fourier transform \cite{8115204}, and data-driven \cite{8839864} operators). One obvious disadvantage lies in the signal-dependent sampling nodes selection and recovery process, which is not suitable for different signal-space caused by perturbations with significantly different spectral characteristics. As such, there is a strong demand to design \textbf{(i) signal-independent} network sampling and recovery schemes, \textbf{(ii) in the absence of explicit dynamic equations}. 

\subsubsection{Koopman Operator}
To address the aforementioned signal-independence and unknown dynamic models, another set of approaches relies on the Koopman operator \cite{williams2015data, 8315070, 8384030, 8431738, 9029917}, which is a linear but infinite dimensional operator that governs the evolution of scalar-value observables (functions) defined on the state space of a nonlinear dynamical system. To approximate a definite-dimensional operator for nonlinear systems with $N$ nodes (variables), extended dynamic mode decomposition (E-DMD) and deep-DMD approaches are developed. For E-DMD, the work in \cite{8431738} developed a polynomial-based Koopman operator for dynamics linearization, by selecting the observables as the $M=O(N^2)$ key polynomial terms of Taylor series (e.g., the multiplicative terms of node 1 and node 2, $x_1\cdot x_2$). 
Similarly, by the multiplications of the Logistic functions defined on each node, they also proposed in \cite{8431525} to generate a group of state-inclusive observables with proved error-bound.
Based on the designed Koopman linearized evolution model, they further derived a minimum number of sampling nodes in \cite{9029917}, by treating the observable set as $\mathbb{R}^M$ and using graph observability analysis  (which maps the sampling nodes to the leading eigenvectors of the Koopman observability gramian).
However, the scheme has two drawbacks. First, to ensure linearization accuracy, the polynomial-based and logistic-based Koopman operator lead to a size explosion ($O(N^2)$) by their multiplicative observables, when addressing large-scale networks (see Figs. \ref{fig5}-\ref{fig6}). Second, the direct use of graph observability analysis  (e.g., rank and gramian analysis) on observable overlooked the intrinsic nonlinear relations between the defined observables, which are all determined by the originally lower sized state space. For example, $x_1$, $x_2$, $x_1x_2$, $x_1^2x_2^2$ are all observables for polynomial-based Koopman linearized model, but are determined by original networked data $x_1$ and $x_2$. Therefore, treating them as $\mathbb{R}^4$ is unreasonable, and will result in extra redundant sampling nodes for signal recovery. We will explain this in greater detail in Section V. A, and in Figs. \ref{fig5}-\ref{fig6}. 

In order to explore a lower-sized Koopman operator, deep-DMD was developed by Yeung, Hodas, and Kundu using Neural networks (NN). The work in \cite{lusch2018deep} further designed an auto-encoder method, by minimizing the mean squared errors (MSEs) of reversible mapping between observables and original states, and of observable and state predictions. The problem lies in that the learned observables may involve coupling signals on different nodes (e.g., one learned observable is $x_2-bx_1^2$ in their discrete spectrum example, containing the signal on node 1 and node 2). This is not suitable for sensor placement, as selecting the leading observables may require to place sensors on every nodes.

\subsection{Contributions \& Organization}
In this work, we propose a novel logarithm-based Koopman and non-linear GFT scheme (abbreviated as Log-Koopman NL-GFT) for sampling and recovering the large-scale networked data. 
The detailed contributions are listed in the following, each addressing an aforementioned shortfall in current approaches:

(1) We propose a logarithm based Koopman operator to linearize the unknown nonlinear networked dynamics. Here, the logarithm-form observables of original state-space are designed to approximate the multi-element multiplicative terms of Taylor series by logarithm summation. In this view, the size of observables can be reduced to $O(N)$, as smaller number of logarithm terms can be used and linearly combined for large number of polynomial-based observables in \cite{8431738}. This suggests the ability of the proposed Log-Koopman to prevent the size explosion when linearizing large-scale networked data. 

(2) We combine the linearization ability of the Log-Koopman operator with a novel nonlinear GFT, by exploiting the nonlinear dependence between the $M$ observables that are defined on the lower size of $N$ original state-space. As such, the proposed Log-Koopman NL-GFT sampling and recovery scheme is able to combine the linear evolution property with nonlinear dependency between observable, thereby outperforming the scheme \cite{8431738} that only relies on graph observability analysis on the Koopman linearized model. Also, other than a signal-space dependent bandlimited property of linear GFT \cite{Chen15, anis2016efficient, Chen16, ortega2018graph, isufi2017observing, 7979500, 7576646, 7934053, 7450694, 7480396, 8115204}, the nonlinear GFT captures the signal-space independent relations of observables, thereby capable of obtaining a signal-independent sampling node set.

(3) We evaluate our proposed Log-Koopman NL-GFT sampling and recovery scheme via two different application domains: (a) networked Biochemical Dynamics of protein-protein interactions, and (b) networked gene Regulatory Dynamics. The results demonstrate that (i) the proposed Log-Koopman operator is able to reduce the observable size to $O(N)$ as opposed to $O(N^2)$ of Poly-Koopman in \cite{8431738}, and (ii) the proposed nonlinear GFT scheme can reduce the number of sampling nodes, compared with the direct use of graph observability analysis in \cite{9029917} after the derivation of Koopman linearized model. This suggests a promising prospect of the proposed Log-Koopman NL-GFT sampling and recovery scheme to a wide range of scientific and engineering monitoring applications.


The rest of this paper is structured as follows. In Section II, we detail the networked nonlinear dynamical system and the problem formulation. In Section III, we provide our designed logarithm based Koopman operator. In Section IV, we elaborate the nonlinear GFT concept and theory. Then, a greedy algorithm for sampling node selection and a gradient descend algorithm for recovery are provided. In Section V, we theoretically compare our proposed Log Koopman NL-GFT scheme with other state of the art approaches. Section VI gives the data-driven experiments and performance discussion. In Section VII, we finally conclude the paper and discuss potential future areas of research.

\section{Model and Problem Formulation}

\subsection{Networked Dynamic Model}
The networked dynamic is described by its underlying graph topology and the dynamic data flow over it. The network topology is configured by a static graph, denoted by $\mathcal{G}(\mathcal{N},\mathbf{A})$. Here, $\mathcal{N}=\{1,\cdots,N\}$ represents a set of node subscripts. $\mathbf{A}$ of size $N\times N$ is the binary adjacent matrix, in which the $(i,j)$th element $a_{i,j}\in\{0,1\}$ reflects the existence of link from node $j$ to node $i$. 

Given the topology of the network, the dynamic data over each node evolves in accordance with its self-dynamic and the coupling interactions from its adjacent nodes. By denoting the data for each $t\in\mathbb{N}^+$ discrete time as a vector of size $N\times1$, i.e., $\mathbf{x}_t=[x_{1,t},\cdots, x_{N,t}]^T$, such evolution can be expressed as:
\begin{equation}
\label{evolution model}
    \mathbf{x}_{t+1}=\mathbf{F}(\mathbf{x}_t, \mathbf{A}),
\end{equation}
where $\mathbf{F}:\mathbb{R}^N\rightarrow\mathbb{R}^N$ is an unknown combined self and coupling nonlinear operator that evolves $t$th state to $(t+1)$th state via the adjacent matrix $\mathbf{A}$. At $t=1$, we regard $\mathbf{x}_1\in\mathbb{R}^N$ as an external input, which is also unknown. 

\begin{figure*}[!t]
\centering
\includegraphics[width=7in]{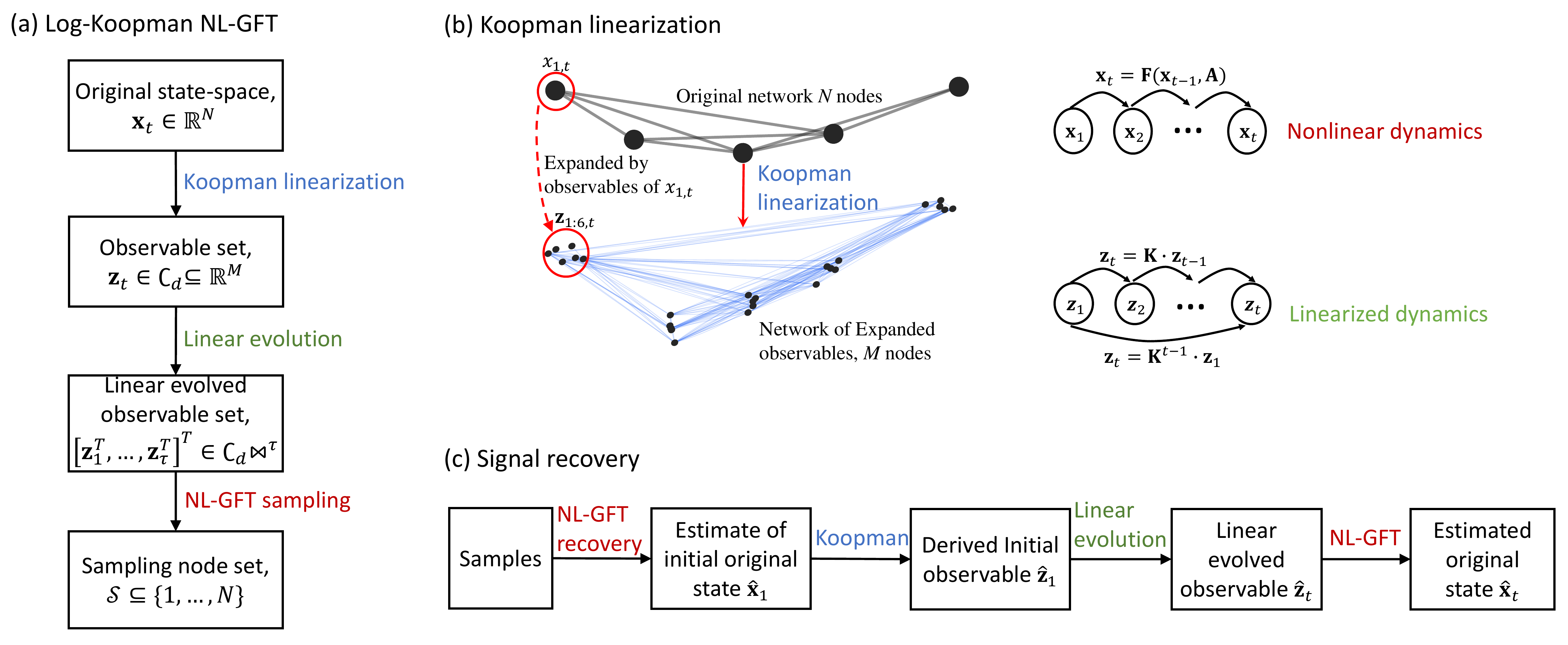}
\caption{Schematic flow of the proposed Log-Koopman NL GFT sampling and recovery method. (a) illustrates the sampling process. (b) shows Koopman linearization, which generates a linear evolution model of extended size $M=O(N)$ observables on original state-space of size $N$. (c) gives the recovery process. }
\label{fig1}
\end{figure*}


\subsection{Problem Formulation}
The purpose of this work is to reconstruct the networked dynamical data via a subset of sampling nodes' data. To be specific, given the sampling node set $\mathcal{S}=\{n_i\}\subset\mathcal{N}$, we define the sampling matrix of size $|\mathcal{S}|\times N$ with elements:
\begin{equation}
\label{sampling_matrix}
    \mathbf{S}=[s_{i,j}],~\text{with~}s_{i,n_i}=1,~s_{i,j\neq n_i}=0.
\end{equation}
Then, the samples collected from the sampling nodes are:
$\mathbf{S}\cdot\mathbf{x}_{t+1}$. 
As such, given the monitoring discrete time-span as $t\in\{1,\cdots,\tau\}$, the aim is to find the sampling node set $\mathcal{S}$ and to design the recovery process to reconstruct $\mathbf{x}_{1:\tau}=[\mathbf{x}_1,\cdots,\mathbf{x}_\tau]$ via the samples $\mathbf{S}\cdot\mathbf{x}_{1:\tau}$.

As aforementioned, the challenges on the design of sampling and recovery methods lie in the absences of both the evolution model $\mathbf{F}$ in Eq. (\ref{evolution model}), and the signal-space (i.e., a subspace of $\mathbb{R}^N$), which make existing works on equation-driven graph observability \cite{isufi2017observing,7852369,zhang2017sensor,7402218,5411741}, and signal-space dependent compression approaches \cite{Chen15, anis2016efficient, Chen16, ortega2018graph, isufi2017observing, 7979500, 7576646, 7934053, 7450694, 7480396, 8115204} less attractive. As such, this motivates our work to 1) approximate a linear evolution model, and 2) find orthogonal nodes for data sampling and recovery.

\subsection{Sketch of Design}
The sketch of the design of sampling and recovery scheme is illustrated via Fig. \ref{fig1}. We firstly adopt the Koopman theory to linearize the unknown nonlinear networked data. Then, the concept and theory of nonlinear GFT will be proposed and used for sampling node selection and signal recovery. We will elaborate them in the following sections.

\section{Koopman Operator and Linearization}
A Koopman operator of one dynamical system is a linear operator that evolves the selected observable functions of the state space as the time advances. By defining the space of all observable functions as $\mathcal{F}$, and stacking such observable functions as $\bm{\psi}=[\psi_1,\cdots,\psi_M]^T$ with $\psi_m\in\mathcal{F}: \mathbb{R}^N\rightarrow\mathbb{R}$ and $m\in\{1,\cdots,M\},M\in\mathbb{N}^+$, the Koopman operator is specified as \cite{williams2015data, 8431738, 9029917}:
\begin{equation}
\label{Koopman model}
    \mathcal{K}\bm{\psi}(\mathbf{x}_t)=\bm{\psi}\left(\mathbf{F}(\mathbf{x}_t)\right)=\bm{\psi}(\mathbf{x}_{t+1}). 
\end{equation}
As such, by selecting appropriate observable functions, one could derive the Koopman operator, and use it as an equivalent evolution model for the corresponding non-linear dynamic system. 


\subsection{State-of-the-art Polynomial-based Koopman Operator}
It is noteworthy that one main difficulty lies in the infinite dimension of $\mathcal{F}$, i.e., $M\rightarrow+\infty$, which makes the Koopman operator $\mathcal{K}$ infinite, and thereby impractical in real-world systems. To address it, many works \cite{williams2015data, 8431738, 9029917} tried to approximate the Koopman operator, by using definite observable functions and span them as the approximated observable space: $\mathcal{F}_D\subset\mathcal{F}$. Specially, the work in \cite{8431738} selected from a proven complete of observable function space leveraged on the polynomial terms of Taylor expansion, i.e., \cite{8431738}
\begin{equation}
\label{o1}
    \mathcal{F}=\left\{\prod_{i=1}^N x_{i,t}^{p_i},~\forall p_i\in\mathbb{N}\right\}.
\end{equation}
By selecting $\bm{\psi}(\mathbf{x}_t)=[x_{i,t}^{p_i}\cdot x_{j,t}^{p_j}]^T$ with $\forall i,j\in\mathcal{N},~p_i,p_j\in\{0,1,2\}$, they constructed the approximated Koopman operator for small-scale (i.e., $N<10$) networked dynamic linearization. However, for large-scale networks ($N>50$), in order to maintain the linearzation accuracy, the scheme leads to a size explosion of observables by selecting complex multi-element multiplicative basic functions (e.g., $x_{i,t}^{p_i}\cdot x_{j,t}^{p_j}\cdot x_{m,t}^{p_m}\cdot x_{n,t}^{p_n}$). We explain this by showing how the multi-element multiplicative terms contributes to the existing observable functions in \cite{8431738}, i.e.,
\begin{equation}
\label{ex1}
\begin{aligned}
    &x_{i,t}\cdot x_{j,t}=f_i(\mathbf{x}_{t-1})\cdot f_j(\mathbf{x}_{t-1})\\
    =&\sum_{\substack{m,n\in\mathcal{N}\\p_i,p_j,p_m,p_n\in\{0,1,2\}}}\alpha_{i,j,m,n}\cdot x_{i,t-1}^{p_i}\cdot x_{j,t-1}^{p_j}\cdot x_{m,t-1}^{p_m}\cdot x_{n,t-1}^{p_n},
\end{aligned}
\end{equation}
where $\alpha_{i,j,m,n}$ denotes the weight. 
As is illustrated in Eq. (\ref{ex1}), the number of 4-element multiplicative terms is increasing with the increase of network scale $N$. Therefore, in order to keep the accuracy of the Koopman linearization, one have to expand the selected observable function set $\bm{\psi}(\mathbf{x}_t)$ that covers such terms. This will lead to a size of $>N^2$ size increase for the approximated Koopman operator, which if used for large-scale network (e.g., $N>50$), may cause heavy computational burden for further sampling selection and signal recovery processes.

\subsection{Logarithm-based Koopman Operator}
To address the aforementioned size explosion, we design a novel group of observable functions that can transform the multiplicative terms (e.g., $x_{i,t}^{p_i}\cdot x_{j,t}^{p_j}$) into summation terms. The idea is the use of logarithm summations to approximate polynomial terms, e.g.,
\begin{equation}
\begin{aligned}
    \log(1+x)+\log(1+y)&=\log\left((1+x)(1+y)\right)\\
    &\approx x+y+xy,
\end{aligned}
\end{equation}
holds, when $x,y\in(0-\delta,0+\delta)$ given $\delta\rightarrow0$. As such, by assigning a constant $C$ such that $sup\{x_{i,t}/C,i\in\mathcal{N},t\in\mathbb{N}^+\}<\delta$, we design the vector-valued observable function as:
\begin{equation}
\label{observables}
    \bm{\psi}\left(\mathbf{x}_t\right)=\left[1, \frac{x_{i,t}}{C}, \log\left(1+\left(\frac{x_{i,t}}{C}\right)^{p_i}\right)\right]^T,~\forall i\in\mathcal{N},
\end{equation}
with some $p_i\in\mathcal{P}\subset\mathbb{N}^+$. Also, we write the vector-valued observable of size $M\times 1$, with its range set $\mathcal{C}_d\subset\mathbb{R}^M$ as:
\begin{equation}
\label{observables1}
    \mathbf{z}_t=\bm{\psi}(\mathbf{x}_t),~\mathbf{z}_t\in\mathcal{C}_d\subset\mathbb{R}^M. 
\end{equation}

Given Eq. (\ref{observables}), we show in the following that each observable function at time $t$ can be evolved and approximated by the summation of others at time $t-1$. The observable function $x_{i,t}/C$ is expressed by:
\begin{equation}
\label{p1}
\begin{aligned}
    \frac{x_{i,t}}{C}&=\frac{f_i(\mathbf{x}_{t-1})}{C}\\
    &=\frac{1}{C}\left(f_i(\mathbf{0})+\mathbf{x}_{t-1}^T\bigtriangledown_{f_i}(\mathbf{0})+\frac{1}{2}\mathbf{x}_{t-1}^T \mathbf{H}_{f_i}(\mathbf{0})\mathbf{x}_{t-1}+o^n\right)\\
    &=\frac{f_i(\mathbf{0})}{C}+\sum_{\substack{m,n\in\mathcal{N}\\p,q\in\mathcal{P}}}a_{m,n,p,q}\cdot\left(\frac{x_{m,t-1}}{C}\right)^{p}\cdot\left(\frac{x_{n,t-1}}{C}\right)^{q}\\
    &\approx\frac{f_i(\mathbf{0})}{C}+\sum_{m\in\mathcal{N}}a_m\frac{x_{m,t-1}}{C}+\sum_{\substack{m\in\mathcal{N}\\p\in\mathcal{P}}}\log\left(1+\left(\frac{x_{m,t-1}}{C}\right)^{p}\right),
\end{aligned}
\end{equation}
where the function $f_i(\cdot)$ is the $i$th element of $\mathbf{F}(\cdot)$, $\bigtriangledown_{f_i}(\cdot)$ is its gradient function, $\mathbf{H}_{f_i}(\cdot)$ is its Hessian matrix. $a_{m,n,p,q}$ and $a_m$ are coefficients invariant with time. The observable function $\log(1+(x_{i,t}/C)^{p_i})$ can be expressed by:
\begin{equation}
\label{p2}
\begin{aligned}
    &\log\left(1+\left(\frac{x_{i,t}}{C}\right)^{p_i}\right)\approx\left(\frac{x_{i,t}}{C}\right)^{p_i}=\left(\frac{f_i(\mathbf{x}_{t-1})}{C}\right)^{p_i}\\
    =&\frac{f_i(\mathbf{0})^{p_i}}{C^{p_i}}+\sum_{ p_1,\cdots,p_N\in\mathcal{P}}b_{p_1,\cdots,p_N}\prod_{ m\in\mathcal{N}}\left(\frac{x_{m,t-1}}{C}\right)^{p_m}\\
    \approx&\frac{f_i(\mathbf{0})^{p_i}}{C^{p_i}}+\sum_{m\in\mathcal{N},p\in\mathcal{P}}b_{m,p}\log\left(1+\left(\frac{x_{m,t-1}}{C}\right)^{p}\right),
\end{aligned}
\end{equation}
where $b_{p_1,\cdots,p_N}$ and $b_{m,p}$ are coefficients invariant with time. 

Given Eqs. (\ref{p1})-(\ref{p2}), any observable functions in Eq. (\ref{observables}) at time $t$ can be approximated from those at time $t-1$, via a linear matrix operator, i.e., the approximated Koopman operator, denoted as $\mathbf{K}$ of size $M\times M$:
\begin{equation}
\label{koopman process}
    \bm{\psi}\left(\mathbf{x}_t\right)=\mathbf{K}\cdot\bm{\psi}\left(\mathbf{x}_{t-1}\right).
\end{equation}
Here, the elements of $\mathbf{K}$ is the coefficients from Eqs. (\ref{p1})-(\ref{p2}). As such, the approximated Koopman operator $\mathbf{K}$ can be derived either from the theoretical deduction if the dynamic model in Eq. (\ref{evolution model}) is known, or from the simulated networked training data. We use the second method in this work, by simulating $D$ groups of training data denoted as $\mathbf{x}_{1:\tau}^{(d)}$ with $d=1,\cdots,D$. Then, by separating the training data into two matrix as $\mathbf{Y}=[\bm{\psi}(\mathbf{x}_{2:\tau}^{1}),\cdots,\bm{\psi}(\mathbf{x}_{2:\tau}^{D})]$, and $\mathbf{X}=[\bm{\psi}(\mathbf{x}_{1:\tau-1}^{1}),\cdots,\bm{\psi}(\mathbf{x}_{1:\tau-1}^{D})]$, we train the Koopman operator $\mathbf{K}$ via:
\begin{equation}
    \mathbf{K}=\argmin\|\mathbf{Y}-\mathbf{K}\mathbf{X}\|_F^2,
\end{equation}
where $\|\cdot\|_F$ denotes the Frobenius norm. The training data are generated via the simulated system with random initialization. For the aim of network sampling, the linearization accuracy of Koopman operator can be further improved by the use of samples. For instance, one can generate new training data with initialization where the values of sampling points are the corresponding initial samples, and others values are random. 

Compared with the existing polynomial-based observable functions for Koopman linearization in \cite{8431738}, the advantage of the proposed logarithm-based observable functions lies in its ability to replace the complex and substantial multi-element multiplicative observable functions with logarithm summation. This reduces the size of the observable $\mathbf{z}_t$ from $O(N^2)$ to $O(N)$ given the comparison from Eq. (\ref{o1}) and Eq. (\ref{observables}), especially for the large-scale network with $N>50$. Also, with the help of the logarithm-form, one observable composition is determined by signals on only one node. This is important for the sensor placement applications, as selecting one observable does not require the placement of sensors on more than one nodes.
As such, we are able to model the unknown evolution of the networked state-space via the linear Koopman operator, which now enables the analysis of optimal sensor placement using linear theory.


\section{Sampling with Nonlinear Graph Fourier Transform}
In this section, we elaborate the concept and sampling theory of the nonlinear GFT. 
The purpose is to determine where to place sensors on nodes of original network in order to recover the initial observable $\hat{\mathbf{z}}_1$, so that the original time-evolved signal states $\mathbf{x}_t$ can be recovered by $\hat{\mathbf{z}}_t=\mathbf{K}^{t-1}\cdot\hat{\mathbf{z}}_1$ and $\hat{\mathbf{x}}_t=\bm{\psi}^{-1}(\hat{\mathbf{z}}_t)$ (shown by Fig. \ref{fig1}(c)).

Notice that, direct use of graph observability methods leads to a larger sampling set than necessary, as they ignore the nonlinear dependency of the initial observable (i.e., treating the elements of $\mathbf{z}_1$ independent). 
Such nonlinear dependency is expressed by the definition of observable function in Eq. (7), i.e., $\mathbf{z}_1=\bm{\psi}(\mathbf{x}_1)$, showing that all $M$ elements of observable $\mathbf{z}_1$ are determined by the $N<M$ independent elements in $\mathbf{x}_1\in\mathbb{R}^N$. 
In this view, we leverage on the observable function $\bm{\psi}^{-1}$ that maps the observable $\mathbf{z}_1$ to its $N$ independent bases, and try to find the sampling set that can recover the signal on $N$ independent bases. This is achieved by the nonlinear GFT concept and sampling theory in the following.

\subsection{Nonlinear GFT Concept}
The general concept of the GFT operator and its bandlimitedness is given as follows:
\begin{Def}
The general GFT operator is an invertible vector-valued function that one-to-one maps a range set $\mathcal{C}_r$ from another set $\mathcal{C}_d$, where $\mathcal{C}_r$ is called the frequency response. We call $\mathcal{C}_r$ a bandlimited frequency response if the size of $\mathbf{x}\in\mathcal{C}_r$ is smaller than that of $\mathbf{z}\in\mathcal{C}_d$.   
\end{Def}

Here, different from the traditional linear GFT where the vector-valued function is a linear operator \cite{Chen15,anis2016efficient,Chen16,ortega2018graph,isufi2017observing,7979500,7576646,7934053,7450694,7480396,8115204}, we generalize the definition which also accounts for the non-linear GFT operator. 

For this work, the Koopman linearization process yields a new network with $M>N$ nodes linked by the Koopman operator $\mathbf{K}$, as is shown in Fig. \ref{fig1}(b). The indexed signals are the $M$ constructed scalar-valued observables in $\mathbf{z}_t=\bm{\psi{(\mathbf{x}_t})}$. As such, a nonlinear GFT operator that combines the network topology and dynamic information can be assigned as the inverse of Koopman observable in Eq. (\ref{observables}), i.e., $\bm{\psi}^{-1}:\mathcal{C}_d\rightarrow\mathcal{C}_r=\mathbb{R}^N$. The frequency response is the original signal with the bandlimitedness property (i.e., $N<M$), which is signal-independent for any $\mathbf{z}_t\in\mathcal{C}_d$.

\subsection{Sampling Theory of Nonlinear GFT}

With the help of the generalized GFT operator, we next propose the nonlinear graph sampling theory, by providing (i) the conditions for the sampling matrix, and (ii) how to recover the signal from the samples.  

\begin{Theo}
\label{t1}
Given a GFT operator $\bm{\psi}^{-1}:\mathcal{C}_d\rightarrow\mathcal{C}_r$, any $\mathbf{z}\in\mathcal{C}_d\subset\mathbb{R}^M$, and a matrix $\bm{\Theta}$ of size $L\times M$, a sampling operator (matrix) $\mathbf{S}_{\Theta}$ of size $S\times L$ ensuring the recovery of $\mathbf{z}$ from $\mathbf{S}_{\Theta}\cdot\bm{\Theta}\cdot\mathbf{z}$ should maintain the one-to-one mapping characteristic of the function $\mathbf{S}_{\Theta}\cdot\bm{\Theta}\circ\bm{\psi}$. The recovered signal of $\mathbf{z}$, denoted as $\hat{\mathbf{z}}$ is expressed as:
\begin{equation}
\label{recover_gft}
    \hat{\mathbf{z}}=\bm{\psi}\left(\left(\mathbf{S}_{\Theta}\cdot\bm{\Theta}\circ\bm{\psi}\right)^{-1}\left(\mathbf{S}_{\Theta}\cdot\bm{\Theta}\cdot\mathbf{z}\right)\right),
\end{equation}
where $\circ$ denotes the function composition operator.
\end{Theo}

\begin{IEEEproof}
We denote frequency response of $\mathbf{z}$ as $\mathbf{x}\in\mathcal{C}_r$. As such, the process of GFT and inverse GFT can be expressed as:
\begin{equation}
\label{gft}
    \mathbf{x}=\bm{\psi}^{-1}(\mathbf{z}),
\end{equation}
\begin{equation}
\label{in_gft}
    \mathbf{z}=\bm{\psi}(\mathbf{x}),
\end{equation}
given the invertible property of the GFT operator. We then multiply the sampling matrix $\mathbf{S}_{\Theta}$ on both side of $\bm{\Theta}\cdot\mathbf{z}=\bm{\Theta}\cdot\bm{\psi}(\mathbf{x})$, i.e., 
\begin{equation}
    \mathbf{S}_{\Theta}\cdot\bm{\Theta}\cdot\bm{\psi}(\mathbf{x})=\left(\mathbf{S}_{\Theta}\cdot\bm{\Theta}\circ\bm{\psi}\right)(\mathbf{x})=\mathbf{S}_{\Theta}\cdot\bm{\Theta}\cdot\mathbf{z}. 
\end{equation}
As such, equation
\begin{equation}
\label{r_fre}
    \mathbf{x}=\left(\mathbf{S}_{\Theta}\cdot\bm{\Theta}\circ\bm{\psi}\right)^{-1}\left(\mathbf{S}_{\Theta}\cdot\bm{\Theta}\cdot\mathbf{z}\right)
\end{equation}
holds if and only if the existence of the inverse function of $\mathbf{S}_{\Theta}\cdot\bm{\Theta}\circ\bm{\psi}$, which is equivalent to its one-to-one mapping characteristic. Then, by taking Eq. (\ref{r_fre}) into Eq. (\ref{in_gft}), the recovered signal $\hat{\mathbf{z}}$ can be computed as Eq. (\ref{recover_gft}). 
\end{IEEEproof}

For the Koopman observable $\mathbf{z}_1$ that has nonlinear dependency i.e., determined by the lower-sized original states, $\mathbf{z}_1=\bm{\psi(\mathbf{x}_1)}$, Theorem \ref{t1} treats it as a bandlimited signal with nonlinear graph frequency response $\mathbf{x}_1$ to the GFT operator assigned as the inverse of Koopman observable function $\bm{\psi}^{-1}$. Then, it proves that the one-to-one mapping from samples to such bandlimited response $\mathbf{x}_1$ can ensure the recovery of $\mathbf{z}_1$. Based on this, following two Propositions are provided to show how such one-to-one mapping can be achieved.

\begin{Prop}
\label{prop1}
Given a GFT operator $\bm{\psi}^{-1}:\mathcal{C}_d\rightarrow\mathcal{C}_r$ with $dim\mathcal{C}$, one prerequisite for signal recovery is that the number of rows of sampling matrix $\mathbf{S}_{\Theta}$ is no lesser than $dim\mathcal{C}$.
\end{Prop}
\begin{IEEEproof}
Otherwise the number of rows of the sampling matrix $\mathbf{S}_{\Theta}$ is $dim\mathcal{C}-1$. Given from Theorem \ref{t1}, $\mathbf{S}_{\Theta}\cdot\bm{\Theta}\circ\bm{\psi}$ is one-to-one mapping. This suggests that all $dim\mathcal{C}-1$ scalar-valued functions of $\mathbf{S}_{\Theta}\cdot\bm{\Theta}\circ\bm{\psi}$ constitute a set of basic functions of $\mathcal{C}_r$, which contradicts its dimension, i.e., $dim\mathcal{C}\neq dim\mathcal{C}-1$. 
\end{IEEEproof}

\begin{Prop}
\label{prop2}
Given a GFT operator $\bm{\psi}^{-1}:\mathcal{C}_d\rightarrow\mathcal{C}_r$ with $\dim\mathcal{C}$, one prerequisite for sampling matrix $\mathbf{S}_{\Theta}$ is that, at least $\dim\mathcal{C}$ scalar-valued functions of $\mathbf{S}_{\Theta}\cdot\bm{\Theta}\circ\bm{\psi}$ are linear independent.
\end{Prop}
\begin{IEEEproof}
Otherwise, if any $\dim\mathcal{C}$ scalar-valued functions of $\mathbf{S}_{\Theta}\cdot\bm{\Theta}\circ\bm{\psi}$ are linear dependent, then there exists $<\dim\mathcal{C}$ linear independent scalar-valued functions that constitute a set of basic function, suggesting $\dim\mathcal{C}_d=\dim\mathcal{C}_r<\dim\mathcal{C}$. 
\end{IEEEproof}

After the elaboration of the non-linear GFT and the rules for the sampling node selection, we next describe how this can be combined with sequential state-space information for optimal sampling and signal recovery.

\subsection{Nonlinear GFT-based Network Sampling and Recovery}
Given the Koopman operator in Eq. (\ref{koopman process}), we specify the Koopman linearized evolution model for discrete time $t=1,\cdots,\tau$ as follows:
\begin{equation}
\label{sequential}
\begin{bmatrix}
\mathbf{z}_1\\\mathbf{z}_2\\\vdots\\\mathbf{z}_\tau
\end{bmatrix}
=
\begin{bmatrix}
\mathbf{K}^0\\\mathbf{K}^1\\\vdots\\\mathbf{K}^{\tau-1}
\end{bmatrix}
\cdot\mathbf{z}_1.
\end{equation}
With the help of the non-linear GFT theory, we assign $\bm{\Theta}=[(\mathbf{K}^0)^T,\cdots(\mathbf{K}^{\tau-1})^T]^T$. The GFT operator is $\bm{\psi}^{-1}$, which maps the range set of the observable $\mathbf{z}_1$ to $\mathbf{x}_1$, the original signals with $dim\mathcal{C}=N$ indexed on $N$ original nodes over graph $\mathcal{G}(\mathcal{N}, \mathbf{A})$. The aim then can be converted to how to determine the sampling nodes set $\mathcal{S}\subset\mathcal{N}$ for the recovery of the frequency response $\mathbf{x}_1$. 

\subsubsection{Selection of Sampling Nodes}
It is noteworthy that the mapping from the sampling node set $\mathcal{S}\subset\mathcal{N}$ of original network, to the sampling matrix $\mathbf{S}_{\Theta}$ in Theorem \ref{t1}, is:
\begin{equation}
\label{relation1}
    \mathcal{S}_{\psi}=\left\{m\Big|\psi_m(\mathbf{x}_t)=\psi_m\left(\mathbf{S}\cdot\mathbf{x}_t\right)\right\},
\end{equation}
\begin{equation}
\label{relation2}
    \mathbf{S}_{\psi}=\left[s_{i,m_i}=1\right],m_i\in\mathcal{S}_{\psi},
\end{equation}
\begin{equation}
\label{relation3}
    \mathbf{S}_{\Theta}=\mathbf{S}_{\psi}\otimes[\underbrace{1,\cdots,1}_{\tau}],
\end{equation}
where $\otimes$ is the Kronecker product. For convenience, we denote the above relations by
\begin{equation}
\label{sampling matrix}
    \mathbf{S}_{\Theta}=\bm{\Gamma}\left(\mathcal{S}\right). 
\end{equation}

Given Theorem \ref{t1}, the optimal selection of $\mathcal{S}$ should ensure the one-to-one mapping characteristic of the function $\mathbf{S}_{\Theta}\cdot\bm{\Theta}\circ\bm{\psi}$, which is a NP-hard challenge. As such, we provide a sub-optimal requirement based on the Propositions \ref{prop1}-\ref{prop2}, aiming to find $\dim\mathcal{C}=N$ linearly independent rows of $\bm{\Theta}$, i.e., 
\begin{equation}
\label{rank_analysis}
    rank(\bm{\Gamma}(\mathcal{S})\cdot\bm{\Theta})=N.
\end{equation}
Here, the difference between Eq. (\ref{rank_analysis}) and the full column-rank sampling selection, i.e., $rank(\bm{\Gamma}(\mathcal{S})\cdot\bm{\Theta})=M>N$ will be detailed in Section V. A.  
We realize Eq. (\ref{rank_analysis}) by minimizing the quotient between the 1st and $N$th singulars of $\mathbf{S}_{\Theta}\cdot\bm{\Theta}=\bm{\Gamma}(\mathcal{S})\cdot\bm{\Theta}$, i.e., 
\begin{equation}
\label{suboptimal}
    \mathcal{S}
    =\argmin_{\mathcal{S}\subset\mathcal{N}}\left\{\frac{\sigma_1\left(\bm{\Gamma}\left(\mathcal{S}\right)\cdot\bm{\Theta}\right)}{\sigma_N\left(\bm{\Gamma}\left(\mathcal{S}\right)\cdot\bm{\Theta}\right)}\right\},
\end{equation}
where $\sigma_i(\cdot)$ denotes the $i$th singular of the matrix. 

Eq. (\ref{suboptimal}) is implemented via a greedy algorithm in Algo. \ref{algo1}. The inputs are the original node set $\mathcal{N}$ from graph $\mathcal{G}(\mathcal{N}, \mathbf{A})$, and the matrix $\bm{\Theta}$ that describes the linear relations between initial observable $\mathbf{z}_1$ and further linear evolved observables $\mathbf{z}_{1:\tau}$. Step 1 is to initialize the sampling node set. Steps 2-5 is to greedily add node with minimum quotient between $1$st and $N$th singulars. The output is the sampling node set $\mathcal{S}$ indicating which nodes are selected for sampling in original graph $\mathcal{G}(\mathcal{N}, \mathbf{A})$.

\begin{algorithm}[t]
\caption{Sampling Node Selection}
\begin{algorithmic}[1]
\Require $\mathcal{N},\bm{\Theta}$
\State Initialize $\mathcal{S}=\emptyset$. 
\While{$\frac{\sigma_1(\bm{\Gamma}(\mathcal{S})\cdot\bm{\Theta})}{\sigma_N(\bm{\Gamma}(\mathcal{S})\cdot\bm{\Theta})}>\gamma$} 
\State $n=\argmin_{n\in\mathcal{N}\setminus\mathcal{S}}\left\{\frac{\sigma_1(\bm{\Gamma}(\mathcal{S}\cup\{n\})\cdot\bm{\Theta})}{\sigma_N(\bm{\Gamma}(\mathcal{S}\cup\{n\})\cdot\bm{\Theta})}\right\}$. 
\State $\mathcal{S}=\mathcal{S}\cup\{n\}$. 
\EndWhile 
\Ensure
Return $\mathcal{S}$.
\end{algorithmic}
\label{algo1}
\end{algorithm}

\subsubsection{Signal Recovery}
With the derivation of the sampling node set $\mathcal{S}$, and its relations to the matrix $\mathbf{S}_{\Theta}$ in Eqs. (\ref{relation1})-(\ref{relation3}), we denote the samples of $[\mathbf{z}_1,\cdots,\mathbf{z}_\tau]$ as:
\begin{equation}
\label{samples}
    \mathbf{y}=\mathbf{S}_{\Theta}\cdot[\mathbf{z}_1,\cdots,\mathbf{z}_\tau]^T. 
\end{equation}
Then, by taking the samples in Eq. (\ref{samples}) into Eq. (\ref{sequential}), and transforming the initial observable $\mathbf{z}_1$ into its graph frequency response, we have:
\begin{equation}
    \mathbf{y}=\mathbf{S}_{\Theta}\cdot\bm{\Theta}\cdot\mathbf{z}_1=\mathbf{S}_{\Theta}\cdot\bm{\Theta}\cdot\bm{\psi}(\mathbf{x}_1). 
\end{equation}
Regarding the difficulty of computing the inverse function $(\mathbf{S}_{\Theta}\cdot\bm{\Theta}\circ\bm{\psi})^{-1}$, we recover the signal $\mathbf{x}_1$ via the quasi-Newton methods\footnote{We use Davidon–Fletcher–Powell (DFP) to approximate and update the inverse of Hessian matrix, and Wolfe conditions for step-size selection.}, by:
\begin{equation}
    \hat{\mathbf{x}}_1=\argmin_{\mathbf{x}_1\in\mathbb{R}^N}\left\{\left\|\mathbf{y}-\mathbf{S}_{\Theta}\cdot\bm{\Theta}\cdot\bm{\psi}(\mathbf{x}_1)\right\|_2^2\right\},
\end{equation}
with gradient:
\begin{equation}
\label{gradient}
    \triangledown=\left(\bm{\psi}(\mathbf{x}_1)^T\cdot\bm{\Theta}^T\cdot\mathbf{S}_{\Theta}^T-\mathbf{y}^T\right)\cdot\mathbf{S}_{\Theta}\cdot\bm{\Theta}\cdot\frac{\partial\bm{\psi}(\mathbf{x}_1)}{\partial\mathbf{x}_1},
\end{equation}
After the computation of $\hat{\mathbf{x}}_1$, we can derive the estimated $\hat{\mathbf{z}}_1=\bm{\psi}(\hat{\mathbf{x}}_1)$, and $\hat{\mathbf{z}}_t=\mathbf{K}^{t-1}\hat{\mathbf{z}}_1$. Then, given the selected observable function in Eq. (\ref{observables}), we finally compute $\hat{\mathbf{x}}_t=\bm{\psi}^{-1}(\hat{\mathbf{z}}_t)$. The process is illustrated in Fig. \ref{fig1}(c).

\section{Novelty Compared with Other State-of-the-Arts}
In this section, we distinguish our proposed Log-Koopman NL-GFT, with other state-of-the-art schemes. 

\begin{figure}[!t]
\centering
\includegraphics[width=3in]{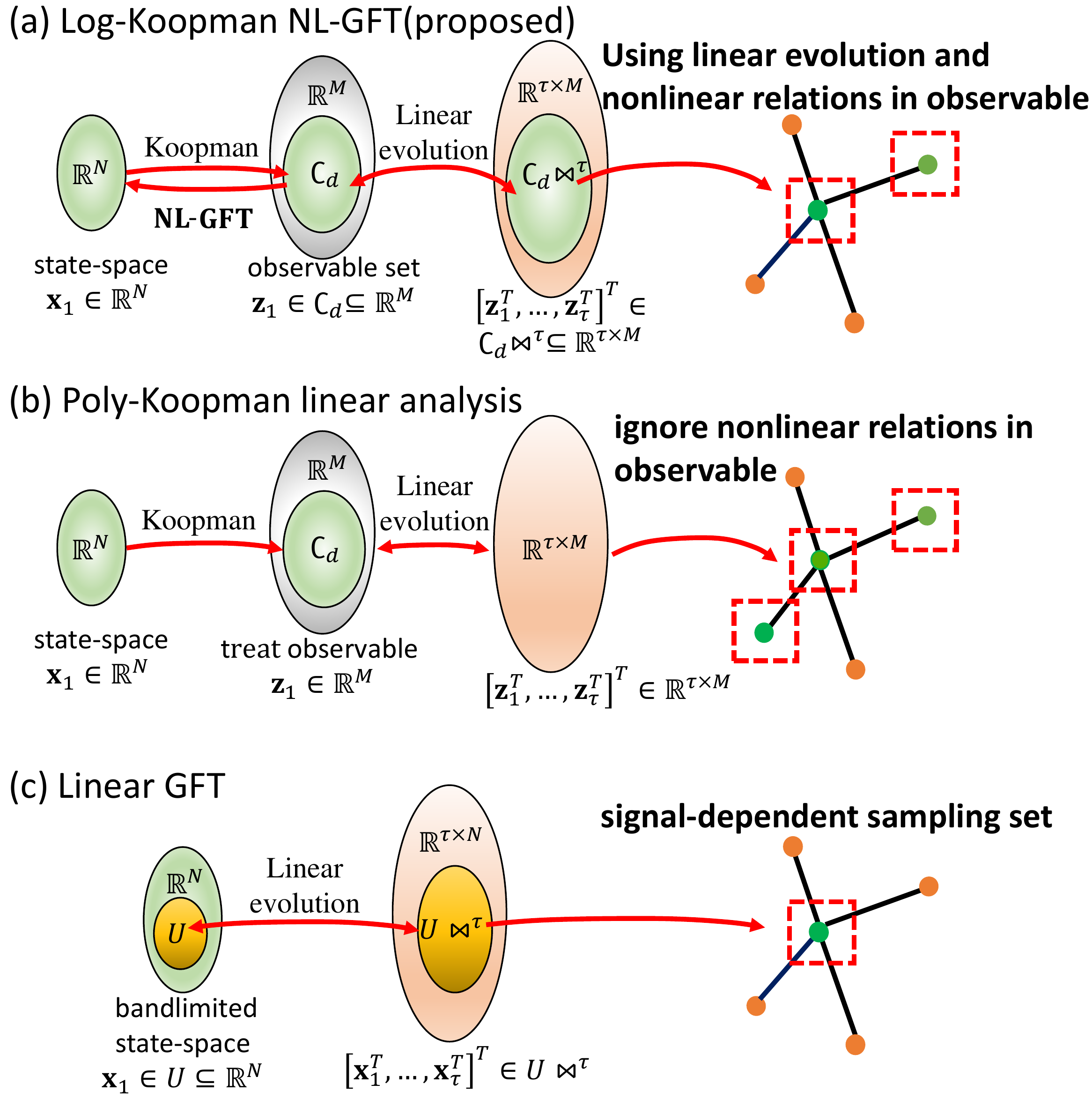}
\caption{Comparison of proposed Log-Koopman NL-GFT method with Poly-Koopman graph observability, and linear GFT method. (a) gives the illustration of proposed Log-Koopman NL-GFT method, whereby the sampling node set maps from the set of NL-GFT signal-independent bandlimited frequency response, i.e.,$\mathbb{R}^N$.
(b) shows the sampling set selection from the Poly-Koopman based graph observability, whereby the initial observable set is treated as a linear space $\mathbb{R}^M$ for graph observability, and therefore maps to redundant sampling node set. (c) shows the linear GFT, whereby the samples are selected for a specific bandlimited initial state-space, suggesting a signal-dependent sampling node selection that is not suitable for signals that are not bandlimited.  }
\label{fig2}
\end{figure}

\subsection{Sampling by Graph Observability Analysis on Poly-Koopman Operator}
After the derivation of Koopman linearized evolution model, i.e., $\mathbf{z}_{t+1}=\mathbf{K}\cdot\mathbf{z}_t$, one straightforward idea is to treat the set of observable as $\mathbb{R}^M$, and use the standard linear theory in \cite{isufi2017observing, 7852369, zhang2017sensor, 7402218, 5411741}, e.g., selecting sampling nodes (corresponding rows) to make $\bm{\Theta}=[(\mathbf{K}^0)^T,\cdots,(\mathbf{K}^{\tau-1})^T]^T$ full column-rank, or by maximizing the state energy computed by Koopman observability gramian. Specially, the latter was implemented by the work in \cite{9029917}, referred to as Poly-Koopman graph observability analysis. The sampling matrix of Poly-Koopman graph observability analysis, denoted as $\mathbf{W}_h$ of size $M\times M$, is derived as \cite{9029917}:
\begin{equation}
\label{reporter}
    \mathbf{W}_h=\left[\mathbf{I}_{k\times k}~\mathbf{0}\right]\cdot\mathbf{V}^{-1},
\end{equation}
by $k$ largest eigenvalues of $\mathbf{K}=\mathbf{V}diag(\lambda_1,\cdots,\lambda_M)\mathbf{V}^{-1}$ to maximize the energy of the observable reports, i.e., 
\begin{equation}
\label{reporter1}
\begin{aligned}
    &\max_{\mathbf{W}_h}\sum_{t=1}^\tau \mathbf{z}_1^T\cdot(\mathbf{K}^t)^T\cdot\mathbf{W}_h^T\cdot\mathbf{W}_h\cdot\mathbf{K}^t\cdot\mathbf{z}_1\\
    =&\sum_{t=1}^\tau \mathbf{z}_1^T\cdot diag\left(\lambda_1^{2t},\cdots,\lambda_L^{2t},0,\cdots,0\right)\cdot\mathbf{z}_1. 
\end{aligned}
\end{equation}

The differences lie in two aspects. 

First, they used the polynomial-based observable function in Eq. (\ref{o1}) to linearize the networked data, which performs accurate linearization approximation for small-scale network. However, when it comes to the large-scale network (e.g., $N>50$), they fall into the size explosion by using $O(N^2)$ terms to construct the observable function $\bm{\psi}(\cdot)$, in order to ensure the linearization accuracy. We explained this by Eq. (\ref{ex1}), and further illustration will be given in Figs. \ref{fig3}-\ref{fig4} and Table I. 

Second, the linear analysis on Koopman linearized evolution model overlooked the nonlinear dependency between elements in the vector-valued observable. This is because both the full column-rank condition of $\bm{\Theta}$, i.e., $rank(\bm{\Gamma}(\mathcal{S})\cdot\bm{\Theta})=M>N$, and the eigenvector analysis in Eq. (\ref{reporter}) treat the initial observable set as the linear space $\mathbb{R}^M$. This therefore overlooks the fact $\mathbf{z}_1\in\mathcal{C}_r\subset\mathbb{R}^M$ with $dim\mathcal{C}_r=N<M$, as the observable $\mathbf{z}_1$ is completely determined by the lower-sized $x_1\in\mathbb{R}^N$, i.e., $\mathbf{z}_1=\bm{\psi}(\mathbf{x}_1)$. As such, the sampling node set maps from $\mathbb{R}^M$ other than $\mathcal{C}_d$ will inevitably result in redundant sampling node for recovering the signal $\mathbf{z}_1$ that belongs to $\mathcal{C}_d$ (seen Fig. \ref{fig2}(a)-(b)). 
We show the comparison performance in Figs. \ref{fig5}-\ref{fig6}.

\subsection{Comparison with Linear GFT Sampling}
Linear GFT sampling method aims at sampling and recovering the networked signal $\mathbf{x}_t$ that belong to a known subspace (also referred to as bandlimited) of $\mathbb{R}^N$, i.e., $\forall t\in\mathbb{N}^+,\mathbf{x}_{t}\in span\{\mathbf{u}_1,\cdots,\mathbf{u}_r\}\subset\mathbb{R}^N$. Here, the orthogonal $r<N$ vectors $\mathbf{u}_1,\cdots,\mathbf{u}_r$ with $r<N$ can be derived either from the $r$-leading eigenvectors of the topology-based Laplacian matrix \cite{Chen15,anis2016efficient,Chen16,7979523, ortega2018graph,isufi2017observing}, or from the simulated data \cite{8839864}. As such, the linear GFT operator $\mathbf{U}^T$ can be assigned as $\mathbf{U}^T=[\mathbf{u}_1,\cdots,\mathbf{u}_r]^T$,
where the processes of GFT and inverse GFT are $\tilde{\mathbf{x}}_t=\mathbf{U}^T\cdot\mathbf{x}_t$ and $\mathbf{x}_t=\mathbf{U}\cdot\tilde{\mathbf{x}}_t$. The sampling matrix $\mathbf{S}_{\Phi}$ to ensure the recovery of $\mathbf{x}_t$ from $\mathbf{S}_{\Phi}\bm{\Phi}\cdot\mathbf{x}_t$ can be determined by \cite{Chen15,anis2016efficient,Chen16,ortega2018graph,isufi2017observing}
\begin{equation}
\label{gft_th2}
    rank\left(\mathbf{S}_{\Phi}\cdot\bm{\Phi}\cdot\mathbf{U}\right)=r,
\end{equation}
where $\bm{\Phi}$ can be a simple identity matrix, or $\bm{\Phi}=[\mathbf{L}^0,\cdots,\mathbf{L}^{T-1}]^T$ in \cite{isufi2017observing} specifies the linear evolved information given $\mathbf{x}_{t+1}=\mathbf{L}\cdot\mathbf{x}_t$. Then, given $\mathbf{S}_{\Phi}$ and the samples $\mathbf{y}=\mathbf{S}_{\Phi}\bm{\Phi}\cdot\mathbf{x}_t$, the recovered signal $\hat{\mathbf{x}}_t$ is \cite{Chen15,anis2016efficient,Chen16,ortega2018graph,isufi2017observing}:
\begin{equation}
    \hat{\mathbf{x}}_t=\mathbf{U}\cdot pinv(\mathbf{S}_{\Phi}\cdot\bm{\Phi}\cdot\mathbf{U})\cdot\mathbf{y},
\end{equation}
where $pinv(\cdot)$ is the pseudo-inverse. 

As is compared by Fig. \ref{fig2}(a) and Fig. \ref{fig2}(c), one difference lies in that the linear GFT sampling method is signal-dependent, since the selection of sampling set is not suitable for the signals that are not belongs to the assumed signal space.  
This leads to the signal-dependent sensor placement, as $\mathbf{S}_{\Phi}$ in Eq. (\ref{gft_th2}) varies with the changes of the assumed signal space. 
Secondly, in the absence of any prior knowledge of the signal model or signal space, their works are unable to generate GFT operator for further network sampling and signal recovery. 
By contrast, our proposed Log-Koopman NL GFT captures the nonlinear bandlimtedness of the observable $\mathbf{z}_t=\bm{\psi}(\mathbf{x}_t)$, which is signal-independent to any vector $\mathbf{z}_t\in\mathcal{C}_d$, therefore leading to a fixed sensor placement scheme for all signals.

\section{Results}
In this section, we evaluate our proposed Log-Koopman NL-GFT sampling method in terms of the sampling rate (i.e., the ratio of number of sampling nodes to total number of network nodes, $|\mathcal{S}|/N$), and the normalized root mean square error (N-RMSE) defined as follows:
\begin{equation}
    \text{N-RMSE}=\sqrt{\frac{\sum_{t=1}^T(\hat{\mathbf{x}}_t-\mathbf{x}_t)^T(\hat{\mathbf{x}}_t-\mathbf{x}_t)}{\sum_{t=1}^T\mathbf{x}_t^T\mathbf{x}_t}}.
\end{equation}

The dynamic network is configured by the Erd\"os–R\'enyi model where each edge is included in the graph with probability $0.5$ that is independent from every other edge. We vary the number of network nodes $N$ from $10$ to $100$, in order to test the applicability of the proposed scheme for various network scales (from small-scale e.g., $N<30$ to large-scale $N\geq50$). Networked data are configured by two general models with rate parameters ($F$, $B$, and $R$) according to \cite{Barzel13}, i.e.,
\begin{equation}
\label{B}
    \frac{dx_i(t)}{dt}=F-B\cdot x_i-\sum_{j=1}^NR\cdot x_i\cdot x_j,
\end{equation}
\begin{equation}
\label{R}
    \frac{dx_i(t)}{dt}=-B\cdot x_i+\sum_{j=1}^NR\cdot \frac{x_j^2}{1+x_j^2}. 
\end{equation}
Eq. (\ref{B}) is referred to as Biochemical Dynamics of protein-protein interactions, in which the rates are configured as $F=10$, $B=1$ and $R=1$, and the initial signal is randomly assigned as $x_i(0)\in(0,1)$ \cite{voit2000computational,Barzel13}. Eq. (\ref{R}) is referred to as gene Regulatory Dynamics, where the rates are set as $B=1$ and $R=1$, and the initial signal is randomly configured as $x_i(0)\in(0,100)$\cite{Barzel13}. Here, it is noteworthy that we do not know the expressions of the models in Eqs. (\ref{B})-(\ref{R}), but only the data generated are used for performance evaluation. For the setting of proposed logarithm observable, by scanning $C$ to obtain best linearization performance, we use $C=500$ in this simulation.

\subsection{Log-Koopman Linearization Performance}
\begin{figure}[!t]
\centering
\includegraphics[width=3in]{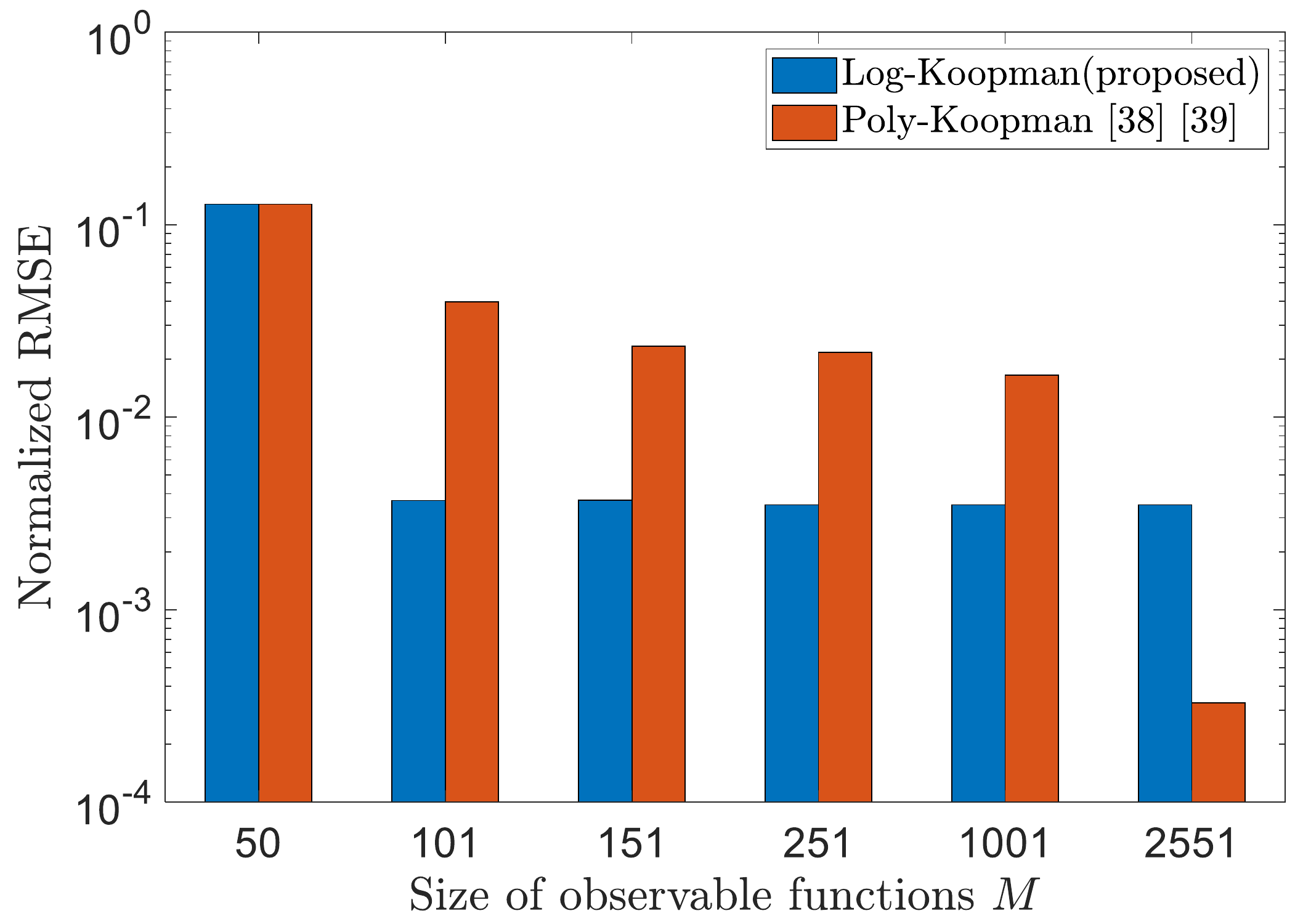}
\caption{Performance comparison between proposed Log-Koopman operator and Poly-based one in \cite{8431738} by $N=50$ networked data of Biochemical Dynamic of protein-protein interactions. The proposed Log-Koopman operator requires less observable functions ($M=O(N)=250$) to approximate the original data, as opposed to the Poly-Koopman operator requiring $O(N^2)=2500$. }
\label{fig3}
\end{figure}

\begin{figure}[!t]
\centering
\includegraphics[width=3in]{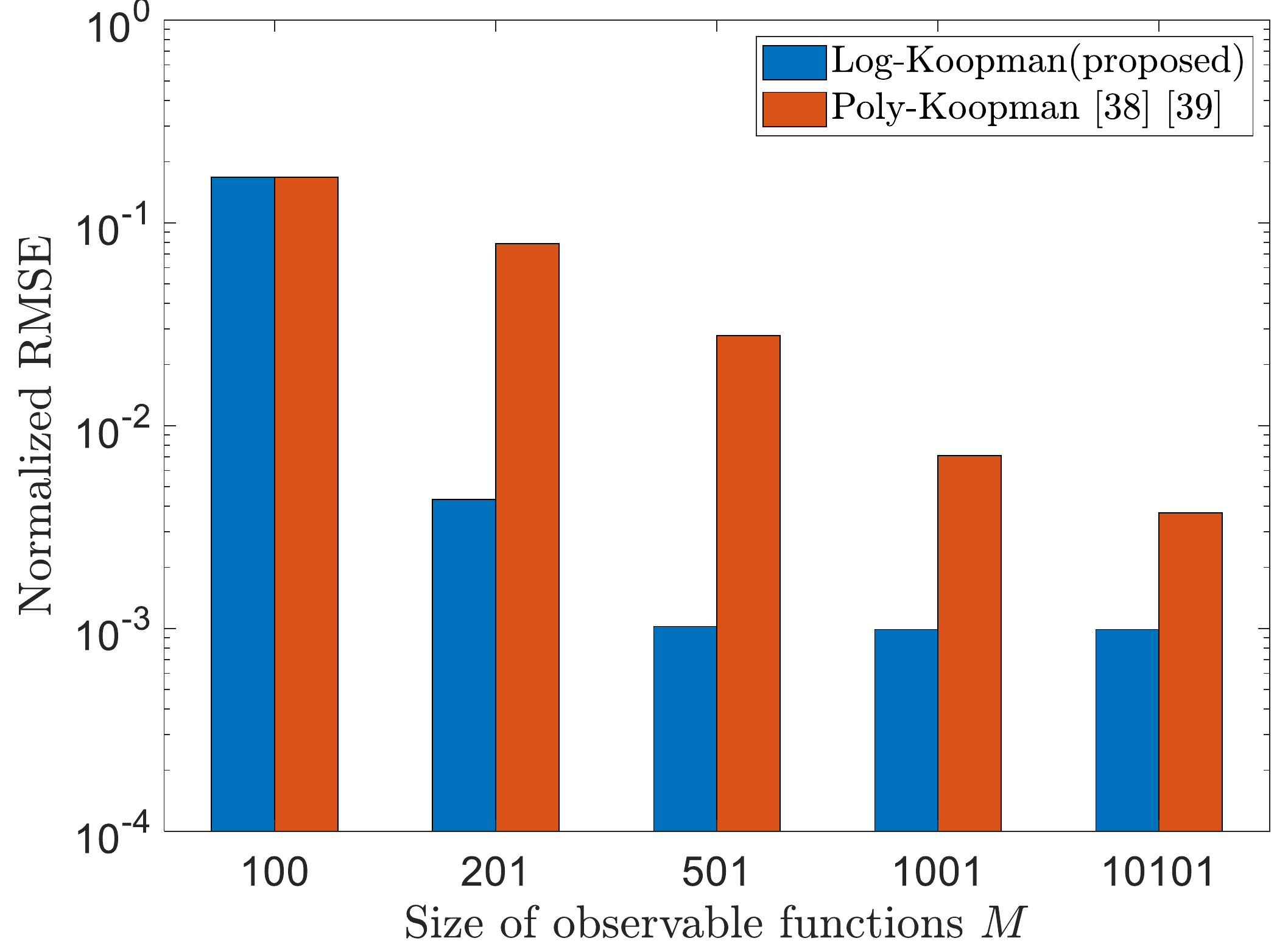}
\caption{Performance comparison between proposed Log-Koopman operator and Poly-based one in \cite{8431738} by $N=100$ gene Regulatory Dynamic networked data. The proposed Log-Koopman operator requires less observable functions ($M=O(N)=10^3$) to approximate the original data, as opposed to the Poly-Koopman operator ($O(N^2)>10^4$).}
\label{fig4}
\end{figure}

We at first test the linearization performance of our proposed logarithm-based Koopman operator in Figs. \ref{fig3}-\ref{fig4}, where the former is for the $N=50$ biochemical network dynamic, and the latter is for the $N=100$ gene Regulatory network Dynamic.
 
The x-coordinate illustrates the number of selected scalar-valued observable functions (i.e., $\psi_1,\cdots,\psi_M$) used for generating the vector-valued observable function $\bm{\psi}(\cdot)$ in Eq. (\ref{Koopman model}). The y-coordinate gives the corresponding N-RMSE between the linerized data and the original data. We can firstly observe that when the observable size equals to the number of network nodes ($M=N$), the normalized RMSEs of two schemes are the same. This is because both the proposed Log-Koopman and the compared Poly-Koopman directly use the original signal state as observable functions (referred to as DMD). Then, with the increase of number of observables, the linearization accuracies of both schemes improve. 

It is seen that the proposed Log-based Koopman operator can reach a small N-RMSE, by using only $M=O(N)$ (e.g., $M=3\times N=150$ in Fig. \ref{fig3}) observable functions. Such number is much lower than that of the Poly-based Koopman operator which requires $O(N^2)$. The reason is attributed to the conversion of the multiplicative terms of Taylor expansion in Eq. (\ref{o1}), to the form of logarithm summations in Eq. (\ref{observables}). As such, only definite number of logarithm-based observable functions are required to approximate and replace the indefinite multiplicative Poly-based observable functions. This suggests the ability of the proposed scheme to prevent the size explosion when linearizing large-scale networked data (i.e., $N>50$), and therefore enables the computational feasibility of further signal processing steps relying on Koopman operator (e.g., the sampling node selection in Algo. \ref{algo1}). 

One drawback of the proposed Log-Koopman operator lies in the existence of N-RMSE lower-bound (e.g., $10^{-3}$ in Figs. \ref{fig3}-\ref{fig4}), due to inaccuracy of the logarithm approximation of the multiplicative polynomial terms. 
We address this by refining the Koopman operator via Eq. (12). Here, the new training data are generated by the initialization where the values of sampling points are the corresponding initial samples, and others values are randomly generated. We will show in the next part that the refined Koopman operator leads to a promising sampling and recovery performances when combined with the nonlinear GFT.

\subsection{Performance of Log-Koopman NL-GFT Sampling and Recovery}
We then evaluate the sampling and recovery performance of the proposed NL-GFT scheme leveraged on the Log-based Koopman operator. 
The compared schemes are the Poly-Koopman based linear state inference in [39], and the batch recovery method using graph smoothness of signal time-variation[28].

\begin{figure}[!t]
\centering
\includegraphics[width=3in]{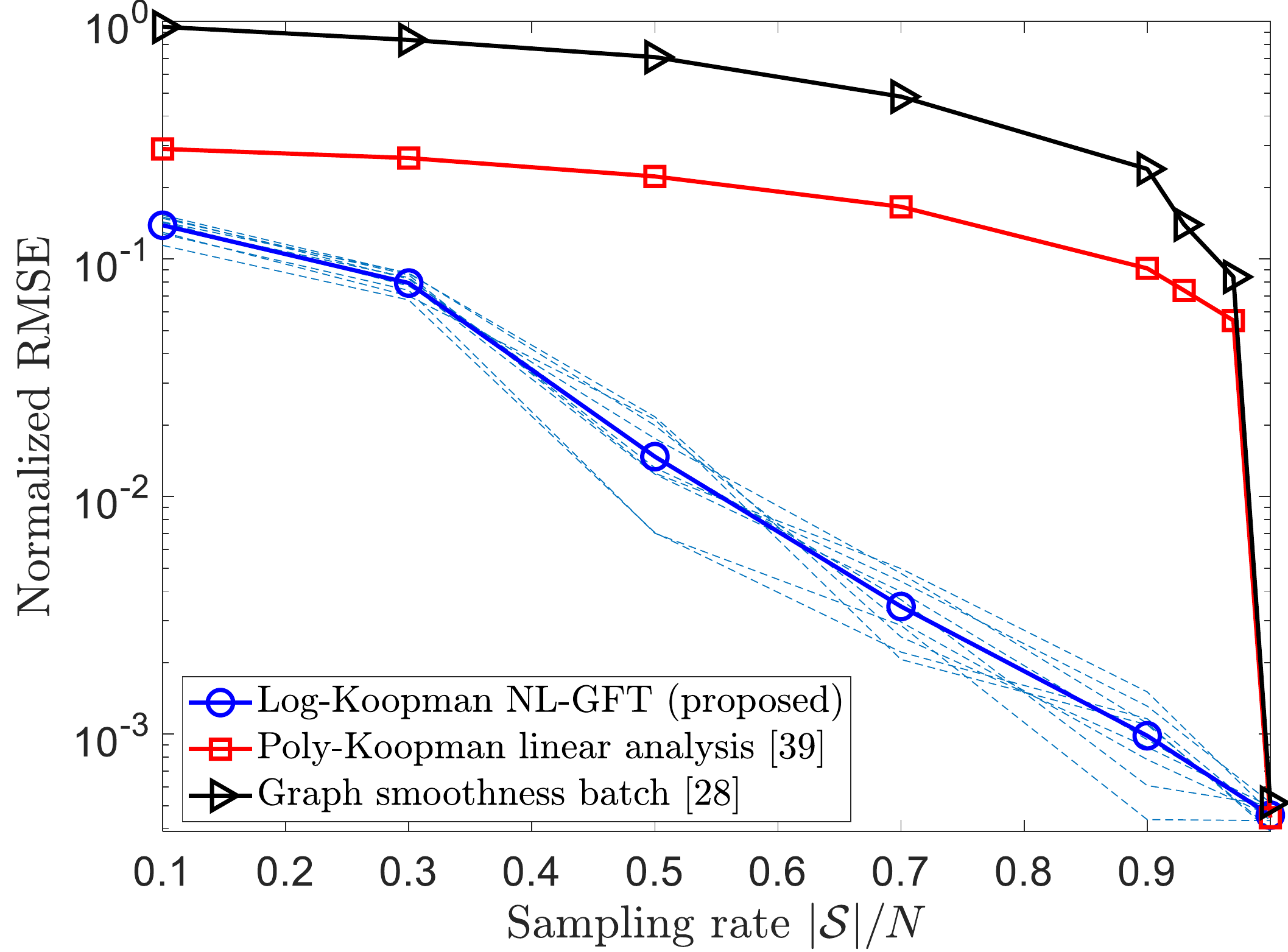}
\caption{Performance comparison between schemes in $N=100$ biochemical network dynamic, where x-coordinate is the sampling rate $|\mathcal{S}|/N$, and y-coordinate is the N-RMSE.}
\label{fig5}
\end{figure}

\begin{figure}[!t]
\centering
\includegraphics[width=3in]{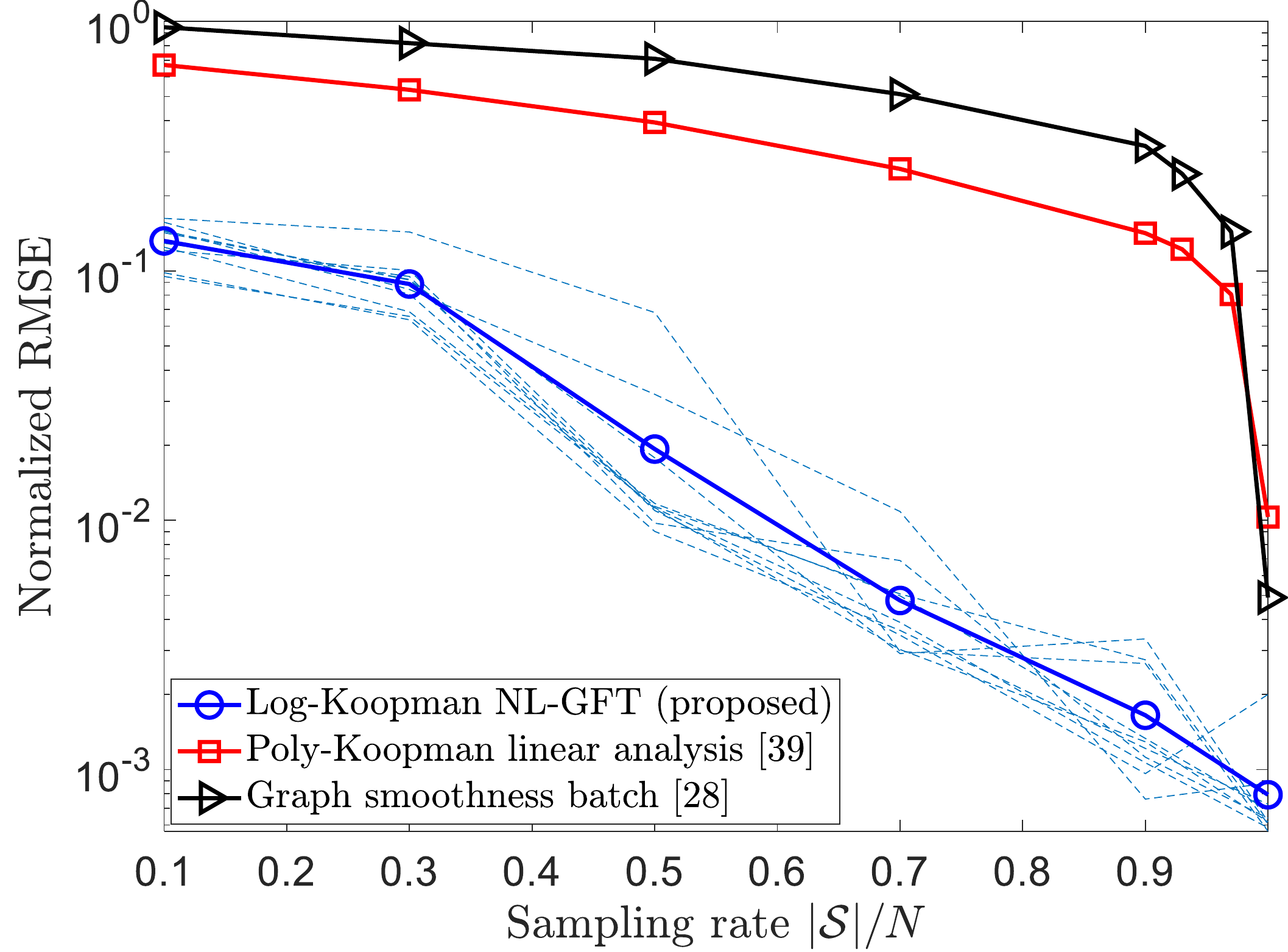}
\caption{Performance comparison between schemes in $N=100$ gene regulatory network dynamic, where x-coordinate is the sampling rate $|\mathcal{S}|/N$, and y-coordinate is the N-RMSE. }
\label{fig6}
\end{figure}

Figs. \ref{fig5}-\ref{fig6} provide the recovery N-RMSE versus the sampling rate, i.e., $|\mathcal{S}|/N$, for $N=100$ biochemical network dynamic, and $N=100$ gene regulatory network dynamic respectively. The dotted lines are the N-RMSE of proposed (blue) schemes for different test data under the same network topology and parameter configuration, and the solid lines give the averages. It is seen from Figs. \ref{fig5}-\ref{fig6} that with the increase of sampling rate, the N-RMSEs of all schemes decreases (e.g., when $|\mathcal{S}|/N\rightarrow1$, all N-RMSEs approach to $10^{-3}$). Then, it is observed that the number of selected nodes from the proposed NL-GFT scheme is much smaller than that of the competitive schemes in [39] and [28]. The proposed scheme can approach an order of $10^{-2}$ N-RMSE by using only half of nodes for sampling, as opposed to the scheme in \cite{9029917} which requires nearly all nodes to ensure the recovery performance. 

We further provide the average recovery N-RMSEs for networks with different scales (from $N=10$ to $N=100$), under various sampling rates $|\mathcal{S}|/N=25\%,50\%,75\%$. Table I is for the biochemical network dynamic, and Table II is for the gene regulatory network dynamic. For each network topology, 1000 data are tested. It is observed that the proposed Log-Koopman NL-GFT scheme has lower average N-RMSE under the same sampling rate, when compared with the Poly-Koopman linear analysis in [39], and with the graph smoothness batch method in [28]. This suggests the general applicability of the proposed scheme for different network scales. Then, it is shown that our proposed scheme requires only half of nodes for sampling to achieve a low N-RMSE as an order of $10^{-2}$, which is smaller than that of the competitive schemes in \cite{9029917} [28] that need almost all the nodes.

We explain such sampling node reduction by two parts. First, when compared with the recovery method using graph smoothness in [28], the proposed Log-Koopman NL-GFT scheme exploits the combined time-evolution and nonlinear dependency, therefore capable of using smaller size of sampling nodes to characterize the whole networked signal.

Second, when compared with the Poly-Koopman linear state analysis method in [39], the sampling node reduction of the proposed scheme is attributed to the use of nonlinear dependency between observables. After the Koopman linearzation, the original $N$ networked data $\mathbf{x}_t$ is expanded by the selected $M>N$ observable functions as $\mathbf{z}_t=[\psi_1(\mathbf{x}_t),\cdots,\psi_M(\mathbf{x}_t)]^T$, and the aim is converted to find the sampling node to recover the initial state $\mathbf{z}_1$ from Eq. (\ref{sequential}). As such, the design of the sampling node selection should take into account the nonlinear dependence between the element of $\mathbf{z}_1$. The graph observability analysis in \cite{9029917} treats the set of observable $\mathbf{z}_1$ as $\mathbb{R}^M$, and ignores such nonlinear dependence, thereby leading to redundant sampling nodes. 
By contrast, our proposed nonlinear GFT sampling method transforms the observable set to its bandlimited set of frequency response, therefore capable of deriving the smaller number of sampling nodes mapping from a lower sized frequency response.

\begin{table}[!t]
\centering
\caption{Biochemical network dynamics, $N=10$ to $100$}
\begin{threeparttable}
\begin{tabular}{llll}
\toprule
 \diagbox{Methods}{N-RMSE}{Sampling rate} & $25\%$  & $50\%$ & $75\%$  \\
 \midrule
 Log-Koopman NL-GFT & 0.1408 & 0.0220 & 0.0022 \\ 
 Poly-Koopman liner analysis [39] & 0.5377 & 0.2501 & 0.2341   \\ 
 Graph smoothness batch [28]. & 0.8831  & 0.7090 & 0.5158 \\
 \hline
\end{tabular}
\end{threeparttable}
\end{table}

\begin{table}[!t]
\centering
\caption{Gene regulatory network dynamics, $N=10$ to $100$}
\begin{threeparttable}
\begin{tabular}{llll}
\toprule
 \diagbox{Methods}{N-RMSE}{Sampling rate} & $25\%$  & $50\%$ & $75\%$  \\
 \midrule
 Log-Koopman NL-GFT & 0.1620 & 0.0338 & 0.0075 \\ 
 Poly-Koopman liner analysis [39] & 0.6474 & 0.4610 & 0.3037   \\ 
 Graph smoothness batch [28]. & 0.8206  & 0.6791 & 0.4766 \\
 \hline
\end{tabular}
\end{threeparttable}
\end{table}

\section{Conclusion}
Networked nonlinear dynamics underpin the complex functionality of many engineering, social, biological, and ecological systems. Monitoring the network's dynamics via subset of nodes is essential for a variety of operational and scientific purposes. For arbitrarily large graphs with nonlinear dynamics, current model-driven methods are dependent on the underlying model assumptions, and data-dependent sampling node selection suffer from either complexity explosion issues or lack of guarantees in performance. One state-of-the-art scheme uses a polynomial based Koopman operator to generate a linear evolution model of observable defined on the original networked state-space, but the sampling node set are still large due to (i) the size explosion of poly-based observables, and (ii) the overlook of nonlinear dependence between observable. 

In this work, we propose a novel logarithm based Koopman operator coupled with a novel nonlinear Graph Fourier Transform (GFT) scheme, entitled as Log-Koopman NL-GFT, for sampling and recovering the networked dynamics. The Log-Koopman operator is able to prevent the size explosion, as logarithm-form observables are designed to replace the substantial multi-element multiplicative poly-observables by logarithm summation. When combined with our novel nonlinear GFT sampling approach, our sampling node set can be completely determined by a bandlimited frequency space \textbf{in a nonlinear manner}. As such, the sampling and recovering algorithms are designed by exploiting the nonlinear dependence of observables. 

The results shows that the proposed Log-Koopman NL-GFT scheme is able to (i) linearize unknown nonlinear dynamics using $O(N)$ observables, and (ii) achieve lower number of sampling nodes, compared with the state-of-the art polynomial Koopman scheme using only the graph observability analysis, which suggests a promising prospect of the proposed Log-Koopman NL-GFT scheme to a wide range of network monitoring applications.

\bibliographystyle{IEEEtran}
\bibliography{main.bib}

\begin{IEEEbiography}[{\includegraphics[width=1in,height=1.25in,clip,keepaspectratio]{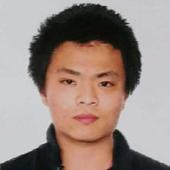}}]{Zhuangkun Wei} received his bachelor's degree, and master's degrees in Electronic Engineering from Beijing University of Posts and Telecommunications (BUPT), Beijing, China in 2014 and 2018 respectively. He is currently pursuing the PhD study at the School of Engineering, The University of Warwick, UK. His research interests cover graph signal processing, molecular communications, and X-AI.
\end{IEEEbiography}

\begin{IEEEbiography}[{\includegraphics[width=1in,height=1.25in,clip,keepaspectratio]{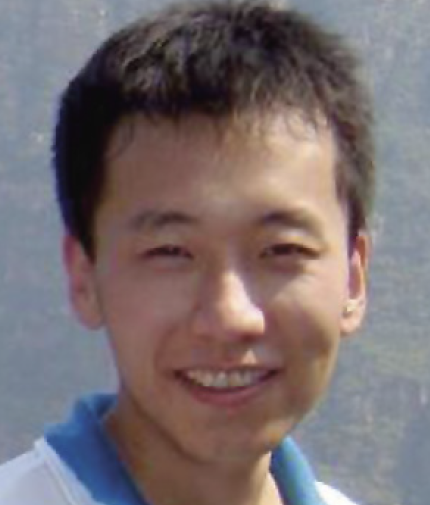}}]{Bin Li} received the Bachelor's degree in electrical information engineering from Beijing University of Chemical Technology (BUCT) in 2007, the Ph.D. degree in communication and information engineering from Beijing University of Posts and Telecommunications (BUPT) in 2013. He joined BUPT at 2013, and now he is an associate professor of the School of Information and Communication Engineering (SCIE). His current research interests are focused on statistical signal processing for wireless communications, e.g. molecular communications, millimeter-wave (mm-Wave) communications and cognitive radios (CRs).
\end{IEEEbiography}

\begin{IEEEbiography}[{\includegraphics[width=1in,height=1.25in,clip,keepaspectratio]{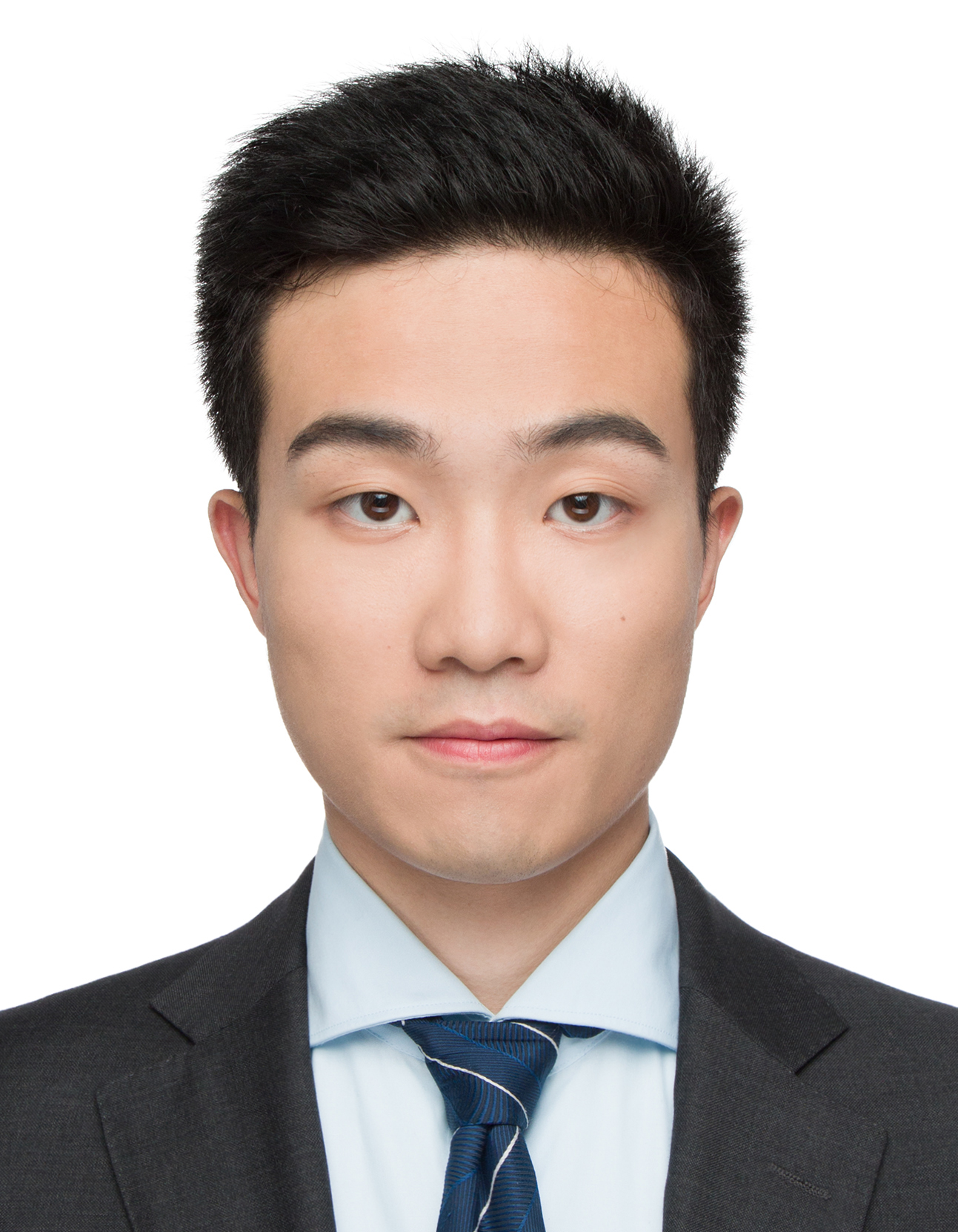}}]{Chengyao Sun} received the BSc degree in applied physics from School of Science at East China University of Science and Technology, Shanghai, China, in 2018. He completed his MSc degree in communications and information engineering at the University of Warwick in 2019, UK. He is currently a PhD student at Cranfield University and his research interests include explainable machine learning algorithms.
\end{IEEEbiography}

\begin{IEEEbiography}[{\includegraphics[width=1in,height=1.25in,clip,keepaspectratio]{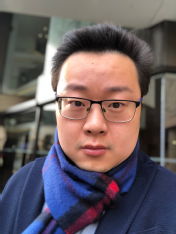}}]{Weisi Guo} (S07, M11, SM17) received his MEng, MA, and Ph.D. degrees from the University of Cambridge. He recently became Chair Professor of Human Machine Intelligence at Cranfield University, and was an Associate Professor at the University of Warwick. He has published over $130$ papers and is PI on over $\pounds2.3$m of research grants from EPSRC, H2020, Royal Society, InnovateUK, and DSTL. His research has won him several international awards (IET Innovation 15, Bell Labs Prize Finalist 14 and Semi-Finalist 16 and 19). He is a Turing Fellow at the Alan Turing Institute.
\end{IEEEbiography}

\end{document}